\documentclass[]{spie}  %>>> use for US letter paper
\usepackage{amsmath,amsfonts,amssymb}
\usepackage{graphicx}
\usepackage[colorlinks=true, allcolors=blue]{hyperref}

\title{The MICADO first light imager for the ELT: the PSF Reconstruction
Software}

\author[a]{Andrea Grazian}
\author[b]{Fernando Pedichini}
\author[a]{Matteo Simioni}
\author[c]{Jani Achren}
\author[a]{Carmelo Arcidiacono}
\author[a]{Luca Cortese}
\author[d]{Ric Davies}
\author[a]{Marco Gullieuszik}
\author[e,f]{Johanna Hartke}
\author[f,e]{Hanindyo Kuncarayakti}
\author[b]{Roberto Piazzesi}
\author[g,h,i]{Elisa Portaluri}
\author[j]{Stefan Raffetseder}
\author[b]{Piero Vaccari}
\author[a]{Benedetta Vulcani}
\author[k]{Roland Wagner}
\author[l]{Anita Zanella}

\affil[a]{INAF-Osservatorio Astronomico di Padova,
  Vicolo dell’Osservatorio 5, I-35122, Padova, Italy}
\affil[b]{INAF - Osservatorio Astronomico di Roma, Via Frascati 33,
  I-00078, Monte Porzio Catone, Italy}
\affil[c]{Incident Angle Oy, Capsiankatu 4A 29, FI-20014, Turku, Finland}
\affil[d]{Max Planck Institut fuer extraterrestrische Physik, D-85748,
  Garching bei Muenchen, Germany}
\affil[e]{Finnish Centre for Astronomy with ESO (FINCA),
  University of Turku, FI-20014, Turku, Finland}
\affil[f]{Tuorla Observatory, Department of Physics and Astronomy,
  University of Turku, FI-20014, Turku, Finland}
\affil[g]{INAF-Osservatorio Astronomico d’Abruzzo,
  Via Mentore Maggini, s.n.c. I-64100, Teramo, Italy}
\affil[h]{INAF - Sezione Universitaria di Lecce, I-73100, Lecce, Italy}
\affil[i]{INFN - Sezione di Lecce, I-73100, Lecce, Italy}
\affil[j]{Industrial Mathematics Institute, Johannes Kepler University Linz,
  Altenberger Strasse 69, A-4040, Linz, Austria}
\affil[k]{RICAM - Johann Radon Institute for Computational and
  Applied Mathematics, Altenberger Strasse 69, A-4040, Linz, Austria}
\affil[l]{INAF - Osservatorio Astronomico di Bologna (OAS), via Gobetti 93/3,
I-40129, Bologna, Italy}

\authorinfo{Further author information: (Send correspondence to A. Grazian)\\
A. Grazian: E-mail: andrea.grazian@inaf.it, Telephone: 0039-049-8293465}

\begin{document} 
\maketitle

\begin{abstract}
The MICADO instrument for the ESO ELT will provide Adaptive Optics
(AO) diffraction-limited NIR imaging and spectroscopy. A critical
component of MICADO is the Point Spread Function Reconstruction (PSF-R)
Software, which is designed to perform a blind reconstruction of the
PSF for all observations using AO telemetry data only, both with the Single
Conjugate (SCAO) and Multi-Conjugate Adaptive Optics (MCAO) modes of
MICADO. The SCAO PSF-R Pipeline is currently being completed, and
validation with ERIS at VLT and SOUL+LUCI at LBT has demonstrated high
accuracy, of percent-level on-axis and 5-20\% off-axis. Initial scientific
simulations confirm that the PSF-R software meets the necessary precision for
high-quality analysis, securing its role for MICADO's first light and
future ELT operations.
\end{abstract}

\keywords{PSF Reconstruction, Post-processing for AO corrected
instrument data, Issues specific to AO for extremely large
telescopes, Modeling, analysis or simulations}

\section{INTRODUCTION}
\label{sec:intro}

The MICADO instrument\cite{Davies16,Sturm24} is designed to provide
diffraction-limited, Adaptive Optics (AO)-assisted Near-Infrared (NIR)
imaging and spectroscopy for the ESO Extremely Large Telescope
(ELT). With its first light just a few years away, MICADO is on
schedule for completion, encompassing both its hardware and software
components. A key element of MICADO is the Point Spread Function
Reconstruction (PSF-R) Software service\cite{Grazian22,Grazian24},
that is an official deliverable of the MICADO project.

The PSF-R tool of MICADO is designed to perform blind reconstruction
of the PSF for all science observations solely from AO telemetry data
in post-processing, without requiring any information from the
scientific data in the focal plane. It will work both with the Single
Conjugate (SCAO) and Multi-Conjugate Adaptive Optics (MCAO) observing
modes of MICADO, the latter allowed by the MORFEO AO
module\cite{Ciliegi24}.
The scientific and technical motivations for a pure PSF reconstruction
tool for MICADO has been summarized in Ref.~\citenum{Simioni22} and
\citenum{Grazian22}.

\section{METHODS}
\label{sec:method}

The method adopted by the MICADO PSF-R tool to reconstruct the PSF
from AO telemetry data only is based on the factorization of the
Optical Transfer Function (OTF). The instrument's OTF terms have been
calculated from the Wavefront Sensor (WFS) and/or Deformable Mirror (DM)
AO telemetry. This technique has been proposed by a seminal idea of
Ref.~\citenum{VeRiMaRo97PSF}, and it has been successfully tested on
PUEO, an AO system at Canada–France–Hawaii Telescope. The PSF-R
software independently calculates various contributions to the
residual phase structure function, distinguishing between frequency
modes corrected by the AO system (OTF parallel) and those that remain
uncorrected (OTF orthogonal or perpendicular). Finally, non-AO
effects, such as the residuals of Non-Common Path Aberrations (NCPA)
and of vibrations, are integrated to obtain the final reconstructed
PSF. The PSF-R Software works both for on-axis\cite{wagner18} and
off-axis\cite{wagner23} observations of MICADO in SCAO mode. It will
also work in the future for MCAO\cite{Wagner22} observations of MICADO
across its entire field of view (FoV), when operated with MORFEO.

\begin{figure}[ht]
\begin{center}
\begin{tabular}{c}
\includegraphics[width=17cm]{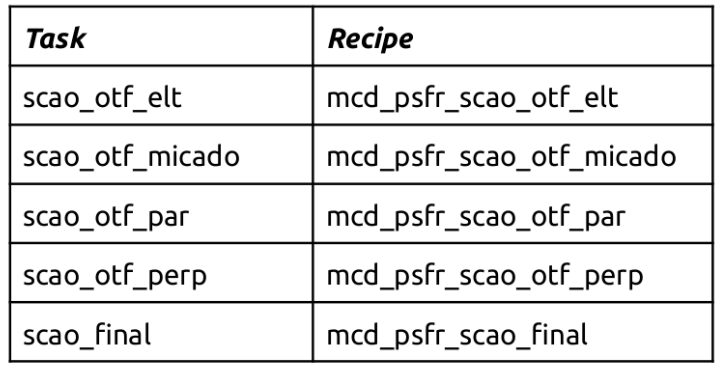}
\end{tabular}
\end{center}
\caption[MICADO PSF-R SCAO Recipes]
{\label{fig:recipe}
The five recipes of the PSF-R SCAO Pipeline of MICADO. The first
recipe is under development, while the last three recipes have been
fully developed and tested both with simulated data of MICADO and real
data from 8-meter class telescopes. The second recipe will be developed
in the near future.}
\end{figure} 

\begin{figure}[ht]
\begin{center}
\begin{tabular}{c}
\includegraphics[width=17cm]{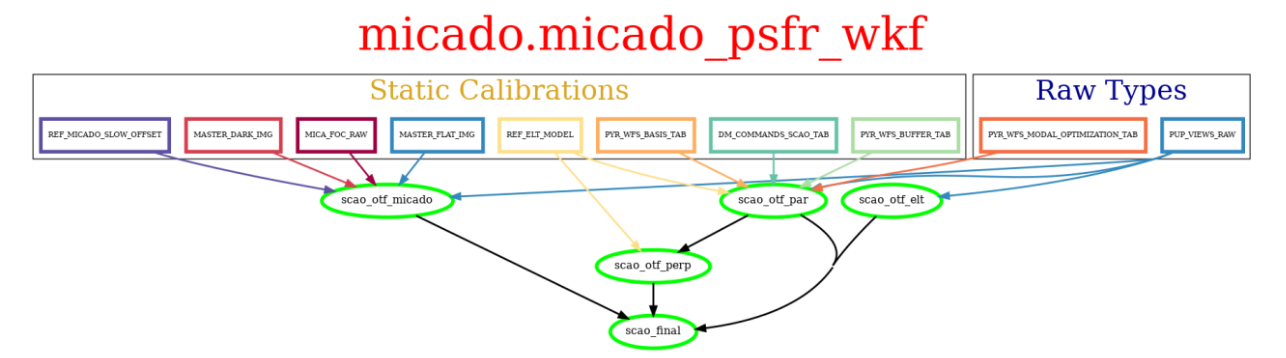}
\end{tabular}
\end{center}
\caption[EDPS Workflow]
{\label{fig:edps}
The EDPS Workflow of the PSF-R SCAO software. Colored boxes indicate
raw input data and calibration files, while green ellipses show the
five recipes. Black arrows indicate data processing, while colored
arrows indicate the link between raw/calibration data and processing
recipes.}
\end{figure}

The PSF-R SCAO software of MICADO has been organized in different main
recipes (see Fig. \ref{fig:recipe}), at a present completion level
(software development) of 70\%. The PSF-R SCAO Pipeline is divided
into five recipes, each one organized into atomic functions with the
aim of classification, organization, and processing of raw and
calibration AO data. These recipes are managed by a high-level tool, the
ESO Data Processing System (EDPS\cite{Freudling24}).
The EDPS Workflow of the PSF-R SCAO software of MICADO is shown in
Fig. \ref{fig:edps}.

Three recipes ($SCAO\_OTF\_Par$, $SCAO\_OTF\_Perp$, $SCAO\_Final$, see
Fig. \ref{fig:recipe}) have been already developed and successfully
run through EDPS, and they have been fully validated with COMPASS
simulated data of MICADO and real data from ERIS\cite{Simioni24} at
VLT and SOUL+LUCI\cite{Simioni22,Simioni22spie,Simioni22sait,Simioni23} at LBT.
Readers interested in this
topic can also see the contribution by Matteo Simioni in this
Conference/Proceeding for the latest news about the validation of the
SCAO PSF-R algorithms. The recipe $SCAO\_OTF\_ELT$ is under
development and it is almost completed, as shown in the contribution
of Jani Achren in this Conference/Proceeding. The development of the
Recipe $SCAO\_OTF\_MICADO$ for non-AO effects of the PSF-R
(e.g. residuals of NCPA and vibrations) is on hold, waiting for real
measurements from MICADO laboratory bench.

\begin{figure}[ht!]
%\begin{center}
\centering  
\begin{tabular}{c}
\includegraphics[width=17cm]{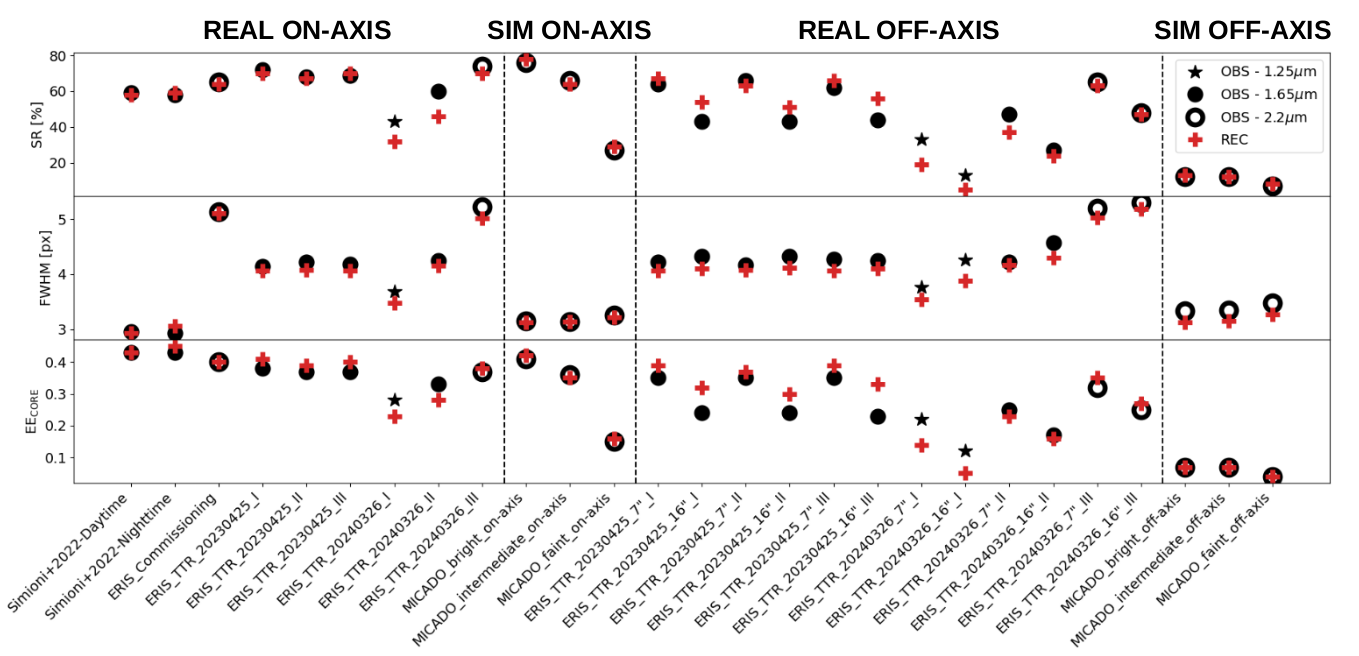}
\end{tabular}
%\end{center}
\caption[PSF-R SCAO Results]
{\label{fig:resu}
Comparison between the observed (black circles) and the reconstructed
(red crosses) PSF parameters (SR, FWHM, EEc) for simulated
(MICADO) and real AO data (from LBT and VLT), both on-axis and off-axis.}
\end{figure}

\section{RESULTS}
\label{sec:result}

The PSF-R SCAO Pipeline has run flawlessly using COMPASS simulated
data under ESO's latest EDPS pipeline framework. The same software has
also been applied to analyze ERIS-VLT AO telemetry data from a bright
three-star asterism\cite{Simioni24}. This validation demonstrated that
the software achieves percent-level accuracy in metrics such as Strehl
Ratio (SR), Full Width at Half Maximum (FWHM), Encircled Energy of the
core (EEc), and half-light radius ($R_{50}$) for the on-axis star.
Validation tests demonstrate an accuracy of 1-5\% for on-axis stars
and 5-20\% for off-axis stars (see Fig. \ref{fig:resu}), confirming that
the PSF-R software meets the requirements for cutting-edge scientific
analysis.

\section{CONCLUSIONS}
\label{sec:conclusion}

No show-stopper for the SCAO PSF-R tool of MICADO has been currently
identified at this advanced level of development. PSF-R algorithms are
going to be completed soon and they will be further improved
thanks to new data from ERIS at VLT and SOUL+LUCI at LBT.
The successful validation of the PSF-R Software of MICADO both with
simulated and observed data firmly places this processing tool on a
trajectory for a highly productive first light of scientific applications
with MICADO\cite{Simioni24eas} and MORFEO\cite{Arcidiacono20,Arcidiacono24}
and regular ELT operations in the near future.

\acknowledgments % equivalent to \section*{ACKNOWLEDGMENTS}
The Italian authors warmly thanks STILES for the generous support
to the MICADO activities.
STILES - STrengthening the Italian Leadership in ELT and SKA is a
program funded by the National Recovery and Resilience Plan (PNRR,
Mission 4, Component 2, Investment 3.1, Project STILES IR0000034 - CUP
C33C22000640006) which aims to strengthen the Italian leadership in the
exploration of the Universe by developing laboratories and instruments
for the two largest ground-based telescopes of the coming decades: the
Extremely Large Telescope (ELT) and the Square Kilometer Array (SKA).
STILES is a program coordinated by the National Institute for
Astrophysics (INAF) in which 7 Italian universities participate, and
it is carried out in collaboration with international research
institutes. The project began in 2023 and officially ended in April
2026.

% References
\bibliography{report} % bibliography data in report.bib
\bibliographystyle{spiebib} % makes bibtex use spiebib.bst

\end{document}